# Quantum Tunneling Enhancement of the C + H$_2$O and C + D$_2$O Reactions at Low Temperature

Kevin M. Hickson,*,[†] Jean-Christophe Loison,[†] Dianailys Nuñez-Reyes[†] and Raphaël Méreau[†]

[†]*Univ. Bordeaux, ISM, CNRS UMR 5255, F-33400 Talence, France*.

**Abstract.**

Recent studies of neutral gas-phase reactions characterized by barriers show that certain complex forming processes involving light atoms are enhanced by quantum mechanical tunneling at low temperature. Here, we performed kinetic experiments on the activated C($^3$P) + H$_2$O reaction, observing a surprising reactivity increase below 100 K, an effect which is only partially reproduced when water is replaced by its deuterated analogue. Product measurements of H- and D-atom formation allowed us to quantify the contribution of complex stabilization to the total rate while confirming the lower tunneling efficiency of deuterium. This result, which is validated through statistical calculations of the intermediate complexes and transition states has important consequences for simulated interstellar water abundances and suggests that tunneling mechanisms could be ubiquitous in cold dense clouds.

**Introduction.**

For many years, activated gas-phase reactions between uncharged species were thought to be unimportant in low temperature environments such as dense interstellar clouds [1]. The rates of such processes were considered to be well described by the Arrhenius expression [2], leading to larger rate constants at higher temperature as the wide distribution of reagent energies allowed the most energetic species to react. At low temperature, the reactivity of these systems was expected to become negligibly small as the reagents were no longer able to surmount the activation barrier; a prediction that was reinforced by experiments above 200 K. Recent work conducted at lower temperatures has challenged this traditional view of reactivity. Hydroxyl radical reactions with several organic molecules [3-7] and the reaction of fluorine atoms with hydrogen [8] have unequivocally demonstrated rate constants that are sometimes enhanced by several orders of magnitude compared with those derived from Arrhenius fits to higher temperature data. One plausible explanation for these dramatic results has been attributed to the possibility of tunneling through the potential energy barrier. Indeed, quantum mechanical theory demonstrates that the wavefunctions describing the reaction intermediates of such complex forming reactions only fall to zero for barriers of infinite height. As the wavefunctions decay exponentially with distance, there is a non-negligible probability that the species can be found on the product side if the barrier is sufficiently thin. For the systems mentioned above, the reagents interact initially through the formation of van der Waal's (prereactive) complexes. In the $OH + CH_3OH$ system for example, ab-initio calculations of the potential energy surfaces (PESs) show that the prereactive complex formed is stabilized with respect to the separated reagents by 20 kJ mol$^{-1}$ [9]. At high temperature, complex lifetimes are generally short due to their high internal energy, so that redissociation to reagents and/or passage over the barrier are the most probable outcomes. At low temperature, the complex lifetime increases and although passage over the barrier is inhibited, exothermic product channels may be accessible by tunneling through the barrier. In the cases mentioned above where enhanced reactivity has been observed, ab-initio calculations show that the most probable reaction pathways at low temperature involve H-atom transfer [3, 5-6, 9-10]. The tunneling efficiency decreases with the particle mass indicating that reactions involving heavier isotopes should be slower, giving rise to noticeable kinetic isotope effects (KIE) such as the one observed for the OH(OD) + CO reaction [11-12]. KIEs have also been observed in the activated $C_2H$ + $CD_4$/$CH_4$ reactions [13], with slower measured rates for the $C_2H$ + $CD_4$ reaction. Here, the difference is adequately explained by the larger separation between the zero-point energy (ZPE) of the reagents and the barrier heights of the respective activated complexes when deuterated methane is employed. As little deviation from Arrhenius behavior is observed for these systems [13], the tunneling contribution appears negligible over the measured temperature range. Interestingly, ab-initio calculations [14] predict that the $C_2H$ + $CD_4$/$CH_4$ reactions do not involve prereactive complex formation, reinforcing the idea that such intermediates could play an important role in raising the tunneling probability [4].

Despite the clear demonstration of large reactivity increases for several gas-phase reactions, there remains an absence of quantitative information regarding the fate of the reagents. One possibility for the processes described above could be complex stabilization rather than reaction, eliminating the need to invoke tunneling effects. Unambiguous product formation has only been demonstrated for the OH + $CH_3OH$ and F + $H_2$ reactions [3, 8], with neither of these studies providing quantitative yields. Such information is crucially important for the accurate prediction of reaction rates in low temperature and low pressure environments such as interstellar space.

Building on these earlier studies of tunneling effects to include isotopic variants of the same molecule, we investigated the reactivity of atomic carbon in its ground electronic state, $C(^3P)$ (hereafter denoted C), with $H_2O$ and $D_2O$ down to 50 K. As gas-phase abundances of C and $H_2O$ are high in dense interstellar clouds [15], this reaction could play an important role in the chemical evolution of such regions. A schematic representation of the PES for the C-$H_2O$ system is shown in Fig. 1, based on current and previous [16-17] ab-initio calculations.

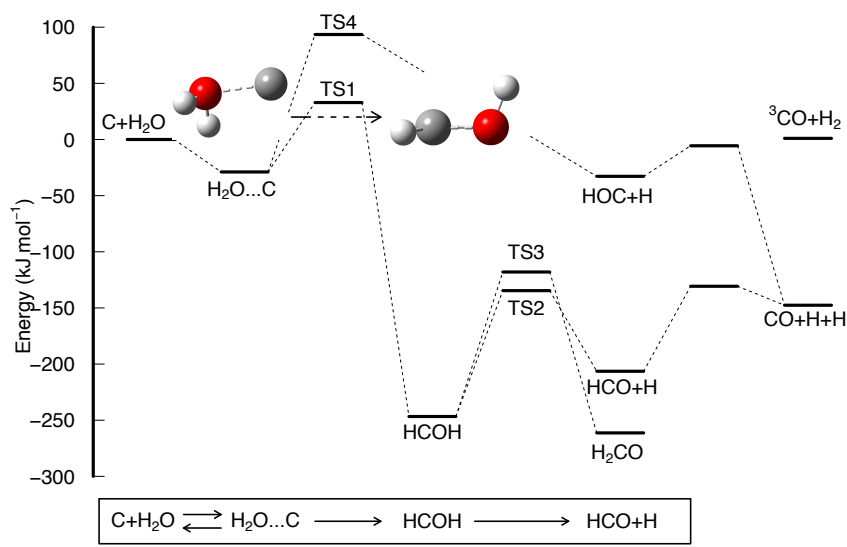

**Figure 1.** Potential energy diagram for the C + $H_2O$ reaction on the triplet surface. Calculations were performed at the CCSD(T)/aug-cc-pVQZ level including ZPE.

Reaction occurs through C-atom addition to the O-atom of water forming a prereactive complex 30-40 kJ mol$^{-1}$ lower than reagents. The probable exothermic products, HCO + H, are separated from the complex by a calculated energy barrier of around 11-40 kJ mol$^{-1}$ [16-18]. Previous room temperature experiments of the gas-phase C + $H_2O$ reaction [19-21] yielded only an upper limit for the rate constant of 1 × 10$^{-12}$ cm$^3$ s$^{-1}$. Consequently, a considerable barrier to product formation is likely, inhibiting reaction at lower temperature in the absence of tunneling effects. Interestingly, earlier low temperature matrix-isolation investigations [18] of this reaction observed the formation of species such as $H_2CO$, although the products were attributed to water reactions with excited state carbon and carbon clusters. To investigate the $C + H_2O/D_2O$ reactions in this work, we employed the CRESU technique [22] (see the Supplementary Information file (SI) for more information). In these experiments, C-atoms were generated by pulsed laser photolysis of carbon tetrabromide ($CBr_4$) and detected by vacuum ultraviolet laser-induced fluorescence (VUV-LIF).

**Results and Discussion.**

Representative decays of the C-atom emission are shown in panels (A) and (B) of Fig. 2 for experiments conducted at 52 K with $H_2O$ and $D_2O$ respectively. An exponential fit to the decays yielded the pseudo-first-order rate constant, $k'$. Without $H_2O$, C-atoms are lost primarily through diffusion from the detection region. However, when excess $H_2O$ or $D_2O$ are added to the flow, C-atom losses are clearly enhanced.

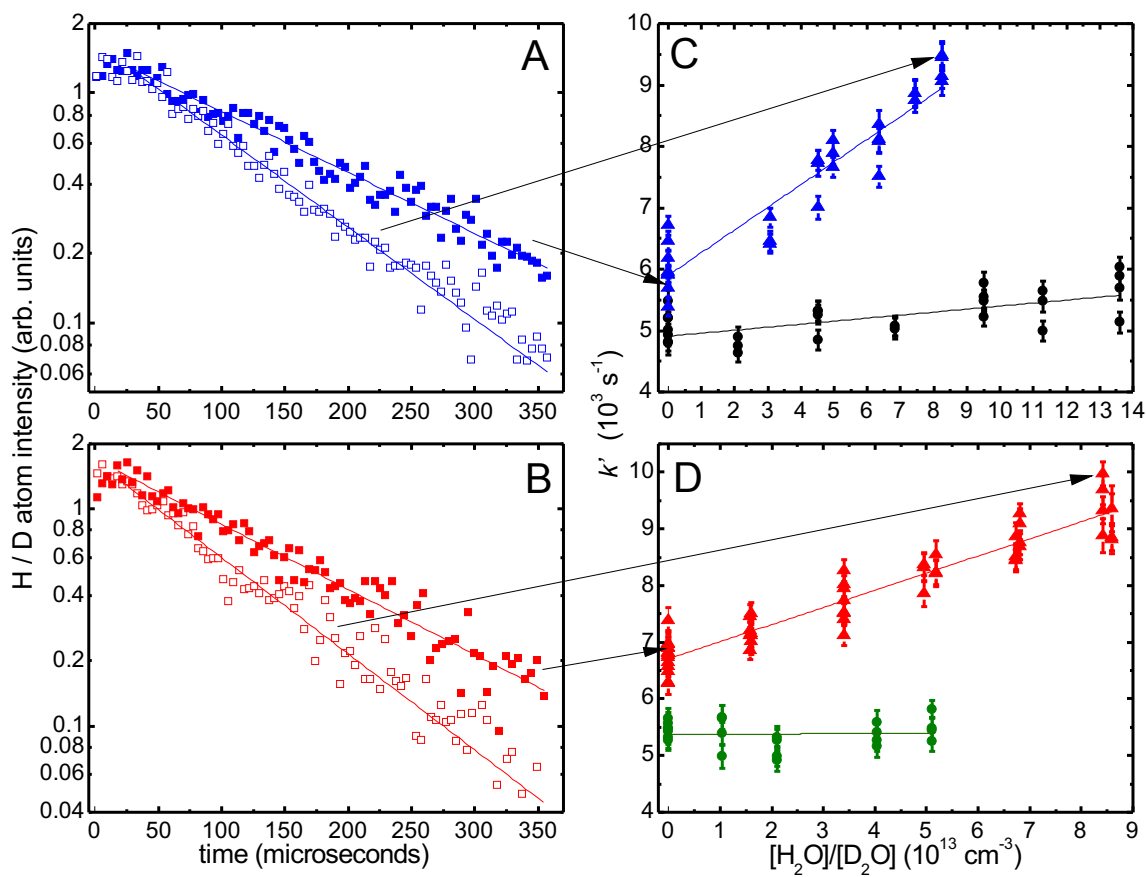

**Figure 2.** Kinetics of atomic carbon removal in the presence of $H_2O$ and $D_2O$. (**A**) Exemplary C-atom temporal profiles at 52 K. (Blue solid squares) without $H_2O$; (Blue open squares) with $[H_2O]$ = 8.3 × $10^{13}$ $cm^{-3}$. (**B**) As (**A**) with data in red and $[D_2O]$ = 8.4 × $10^{13}$ $cm^{-3}$. Lines represent exponential fits to individual decays. (**C**) Pseudo-first-order rate constants for the C + $H_2O$ reaction as a function of $[H_2O]$ at 106 K (black solid circles) and 52 K (blue solid triangles). (**D**) As (**C**) for the C + $D_2O$ reaction at 106 K (green solid circles) and 52 K (red solid triangles). Solid lines represent weighted linear-least squares fits to the data yielding second-order rate constants. The error bars reflect the 1σ uncertainties obtained by fitting to C-atom VUV LIF profiles such as those shown in panels (**A**) and (**B**).

Rate constants were determined from a weighted linear least-squares analysis of plots of the individual $k'$ values versus [$H_2O$] or [$D_2O$]. Panels C and D of Fig. 2 show the second-order plots for both reactions at 52 K and 106 K.

The measured second-order rate constants are summarized in Table S1. At 106 K, the reaction with deuterated water was effectively too slow to be measured and experiments performed above 106 K could not identify any supplementary C-atom losses for either system. Our experimental results are compared to previous work in Fig. 3.

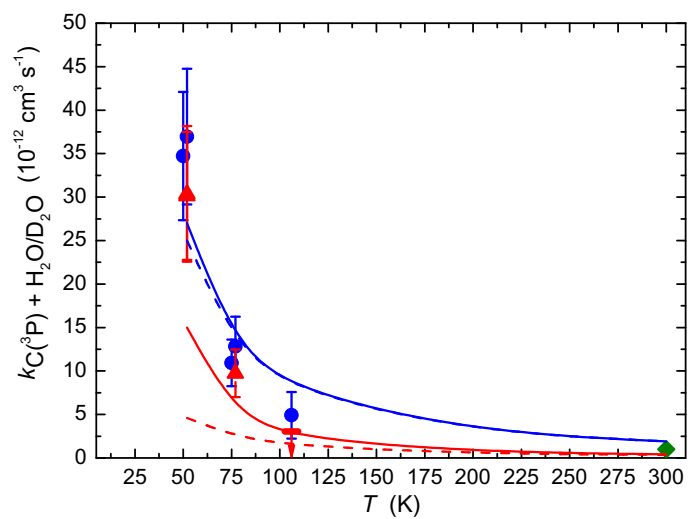

**Figure 3.** Rate constants as a function of temperature. The C + $H_2O$ reaction: (green diamond) Husain and Kirsch [19]; (solid blue circles) these experiments; (solid blue line) calculated total rate constant (stabilization + reaction); (dashed blue line) the reactive rate constant. The red symbols and lines represent the equivalent quantities for the C + $D_2O$ reaction. Error bars on the present values indicate the combined statistical uncertainty (1σ) and an estimated systematic uncertainty.

The rate constants for both reactions increase dramatically below 100 K, with higher values for the C-$H_2O$ system. At this stage it is impossible to confirm whether these rates correspond to reaction through tunneling, complex formation, or a combination of the two effects. Indeed, carrier gas collisions could induce C..OH$_2$(OD$_2$) prereactive complex stabilization [12]; an outcome which would be indistinguishable from reaction by following C-atom losses alone. To discriminate between reactive and non-reactive losses, we examined the product channels for these reactions by following H- or D-atom formation directly by VUV-LIF. To obtain quantitative product yields, the H-atom intensities were compared with H-atom signals generated by the C + C$_2$H$_4$ reaction (with a H-atom yield close to unity [23]). As the D and H line strength factors are essentially identical [24], it was also possible to obtain quantitative D-atom yields using the same reference by maintaining a fixed laser energy to probe each atomic transition. Typical H-/D-atom formation curves are presented in Fig. 4 for the three reactions at 52 K.

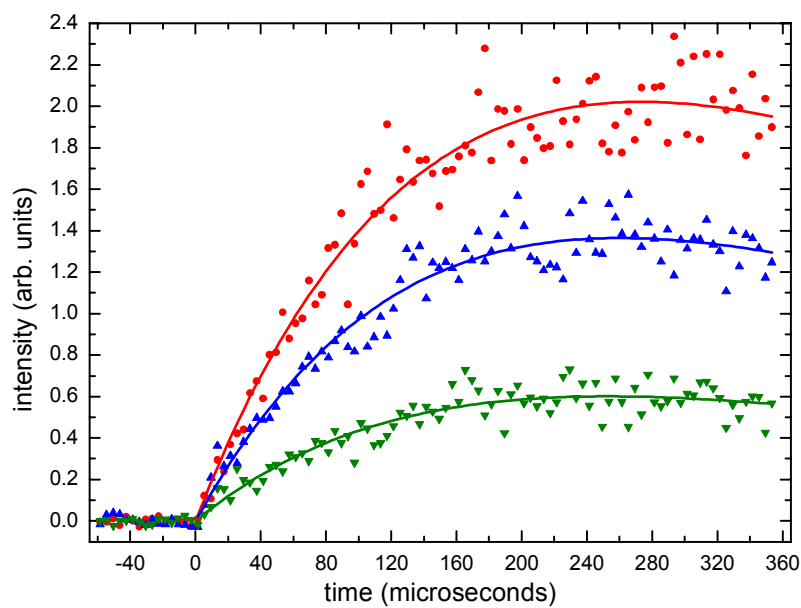

**Figure 4.** Relative yields of atomic hydrogen and atomic deuterium products. H/D-atom temporal fluorescence profiles at 52 K. (Red circles) H-atoms from the C + C$_2$H$_4$ reaction ([C$_2$H$_4$] = 1.5 × 10$^{12}$ cm$^{-3}$); (Blue triangles) H-atoms from the C + H$_2$O reaction ([H$_2$O] = 1.8 × 10$^{13}$ cm$^{-3}$); (Green triangles) D-atoms from the C + D$_2$O reaction ([D$_2$O] = 2.0 × 10$^{13}$ cm$^{-3}$). Coreagent concentrations

were chosen to yield similar first-order production rates for all reactions considering the respective rate constants obtained by detecting C-atoms.

H-/D-atoms are clearly detected, proving that tunneling is an important factor for both reactions. No H-/D-atoms were observed in experiments performed above 106 K, indicating the negligible role of tunneling at higher temperatures; in good agreement with the low rate constants obtained through C-atom kinetic measurements. The H- and D-atom yields from the $C(^3P) + H_2O/D_2O$ reactions are substantially smaller than the corresponding reference one, with values of (0.58 ± 0.04) and (0.29 ± 0.02) respectively at 52 K indicating that large fractions of $C..OH_2(OD_2)$ complexes do not lead to reaction. These values assume that energized HCO/DCO coproducts do not fragment, as HCO/DCO dissociation would lead to secondary H-/D-atom formation, further decreasing the measured product yields.

To reconcile the results of kinetic and product studies, and to provide a basis to predict rate constants under interstellar conditions, we performed statistical calculations of the $C-H_2O/D_2O$ systems employing the Master Equation Solver for Multi Energy Well Reactions (MESMER) [25]. The experimental observations (rate constants and product branching ratios) were used to constrain the initial parameters of the master equation. It was necessary to adjust the complex well depth and barrier width to reproduce the experimental results (see SI file for more details), although these modifications were consistent with those used in previous calculations of similar systems [3, 6, 26]. These calculations, shown in Fig. 3 capture the main experimental features; (I) a rapid rise in the total rate constant for both systems below 100 K (II) a substantially larger fraction of complex stabilization to the total rate for the $C + D_2O$ reaction, although the reactive contribution for the $C + H_2O$ system is somewhat overestimated.

Ground state atomic carbon has been observed in dense interstellar clouds through its ground state fine structure line emission at 492 GHz [27], indicating that it could be one of the most abundant species in these regions, with current models [15] adopting peak gas-phase C-atom abundances of $10^{-4} - 10^{-5}$ relative to $H_2$. Observations of $H_2O$ in dark clouds such as TMC-1 and L134N typically lead to abundances in the $10^{-7} - 10^{-8}$ range. To determine the eventual role of the $C + H_2O$ reaction in such environments, we used the statistical calculations to extend the measured rates to representative interstellar cloud temperatures and pressures, yielding a reactive rate constant of $2 \times 10^{-10}$ cm$^3$ s$^{-1}$ at 10 K in the low pressure limit. To estimate its relative importance with respect to other reactions involving water already present in astrochemical networks, we used the reactive flux as a guide (the product of the individual abundances and the rate constant); a comparison which confirms that the $C + H_2O$ reaction is the main sink for gas-phase water in dense interstellar clouds. Indeed, the reactive flux is several orders of magnitude larger than predicted for other systems where tunneling has been invoked, such as the OH + methanol reaction [3]. While this is an important conclusion in itself, as astrochemical simulations overestimate water abundances by at least an order of magnitude [15], the major interest of this reaction could be its influence on interstellar dust grain chemistry. Such particles are surrounded by a water-ice mantle containing non-negligible quantities of $CO_2$, CO and $CH_3OH$ with traces of other carbon, nitrogen and sulphur containing molecules [28]. Astrochemical models predict that gas-phase C-atoms deplete onto these ices, reacting predominantly with H-atoms to form $CH_4$ [29]. Introducing the $C + H_2O$ reaction into models provides an alternative pathway for carbon loss, to yield one or both association products HCOH and $H_2CO$. As two hydrogenation steps of the standard $CH_3OH$ formation mechanism (H + CO → HCO and H + $H_2CO$ → $H_2COH/CH_3O$) are characterized by activation barriers [30], the C

+ H$_2$O reaction could facilitate methanol formation through the barrierless H + HCOH reaction if HCOH→H$_2$CO isomerization is inefficient [31].

Previously, activated reactions were considered unimportant at low temperature, and such processes have mostly been discarded from reaction networks. This and other recent work demonstrate that many processes characterized by barriers could be enhanced by tunneling and that their relevance to models of cold environments needs to be reevaluated.

**Experimental and Theoretical Methods.**

**Experiments.** The continuous flow CRESU apparatus used in the present work has been described previously [32-33]. Four Laval nozzles were used in this study whose supersonic flow characteristics (temperatures, densities and velocities) are listed in Table S2. C($^3$P) atoms were generated by the multiphoton dissociation of CBr$_4$ molecules at 266 nm by aligning the photolysis laser along the supersonic flow axis. Previous studies [32] under similar conditions have shown that excited state carbon atoms C($^1$D) are also produced, at the level of 10-15% with respect to C($^3$P). Rate constants for the C($^3$P) + H$_2$O reaction were measured by detecting C($^3$P) atoms directly by on-resonance VUV LIF at 127.755 nm. In later product formation measurements, H($^2$S) or D($^2$S) atoms were detected through VUV LIF at 121.567 nm and 121.534 nm respectively.

H$_2$O was introduced into the gas flow upstream of the Laval nozzle using a controlled evaporation mixing (CEM) system. Flows of liquid water were passed into an evaporation device heated to 373 K along with a small flow of Ar or N$_2$ to carry water vapour into the reactor. The gas-phase water concentration was determined by absorption using the 185 nm line of a mercury lamp upstream of the reactor. D$_2$O could not be introduced into the experiments in this way, given the need for a large volume of D$_2$O in the CEM reservoir. Moreover, the absorption cross-section of D$_2$O is almost an order of magnitude smaller than the $H_2O$ one at 185 nm, so it was no longer possible to determine its concentration spectroscopically. Instead, $D_2O$ vapour was carried into the reactor using a room temperature bubbler connected to a cold trap held at 288 K to precisely control the concentration of $D_2O$ in the flow [34]. The pseudo-first-order approximation was used in all experiments with $[H_2O]/[D_2O] \gg C(^3P)$ so that the temporal dependence of the $C(^3P)$ LIF signal was given by a simple exponential function of the form $A \times e^{-Bt}$. For experiments monitoring the formation of product H/D-atoms, a biexponential function of the form $A \times (e^{-Bt} - e^{-Ct})$ was used to describe the initial production and eventual loss of these atoms.

**Calculations.** Rate constants and product branching ratios were calculated using the Master Equation solver MESMER [25], employing a new PES calculated here (see SI file for more details). Briefly, the phase space for all wells on the PES was divided into grains of a set size and isoenergetic, microcanonical rate coefficients for reactive processes between different species were calculated using Rice, Ramsperger, Kassel and Marcus (RRKM) theory. Energy transfer probabilities between different grains of the same species were obtained using an exponential down model [35]. For all species, external rotations were considered to be rigid rotors whilst vibrational modes were treated as harmonic oscillators. For the current calculations, a grain size of 20 $cm^{-1}$ was used above 150 K although this was reduced to 10 $cm^{-1}$ at the lowest temperatures. Harmonic frequencies and rotational constants were taken from CCSD(T) calculations, but given the large range of energies for the stationary points between the different levels of theory, the energy of the complex was varied during the calculations as well as the width of the barrier connecting the complex and the HCOH adduct. During the procedure to reproduce the experimental data, the same adjustments were applied to both the C + $H_2O$ and C + $D_2O$ systems for the energies of the prereactive complexes as well as to the width of the barrier to maintain a consistent description of these two reactions. To account for quantum mechanical tunnelling, the tunnelling transmission coefficients were calculated using the semi-classical WKB method [36]. With this method, considered to be more accurate than the one based on a parabolic Eckart-type barrier, the transmission coefficients are calculated through a one dimensional potential employing the mass-weighted intrinsic reaction coordinate (IRC) calculated at the MP2 level. Nevertheless, as this approach may underestimate the tunneling contribution as curvature could lead to an overly long tunneling path, it was necessary to reduce the barrier width to reproduce the experimental results. The barrierless complex formation step was treated using the inverse Laplace transform method to convert the high pressure limiting canonical rate coefficients for the initial association in Arrhenius form into microcanonical rate coefficients. More detailed information on the experimental and theoretical methods is given in the SI file.

**Author information**

**Corresponding author**

*E-mail: kevin.hickson@u-bordeaux.fr

**Acknowledgments**.

The authors acknowledge support from the Agence Nationale de la Recherche (ANR-15-CE29-0017-01) and from the French INSU/CNRS programs 'Physique et Chimie du Milieu Interstellaire' (PCMI) and 'Programme National de Planétologie' (PNP).

**Supporting Information Available:**

The supporting information file includes details of the experimental and theoretical methods used, supplementary figures S1 to S7 and supplementary tables S1 to S5.

**References and notes.**

1.  Viala, Y. P. Chemical Equilibrium from Diffuse to Dense Interstellar Clouds .1. Galactic Molecular Clouds. *Astron. Astrophys. Supp. Series* **1986,** *64*, 391-437.

2. Smith, I. W. M. The Temperature-Dependence of Elementary Reaction Rates: Beyond Arrhenius. *Chem. Soc. Rev.* **2008,** *37*, 812-826.
3. Shannon, R. J.; Blitz, M. A.; Goddard, A.; Heard, D. E. Accelerated Chemistry in the Reaction Between the Hydroxyl Radical and Methanol at Interstellar Temperatures Facilitated by Tunnelling. *Nat. Chem.* **2013,** *5*, 745-9.
4. Caravan, R. L.; Shannon, R. J.; Lewis, T.; Blitz, M. A.; Heard, D. E. Measurements of Rate Coefficients for Reactions of OH with Ethanol and Propan-2-ol at Very Low Temperatures. *J. Phys. Chem. A* **2015,** *119*, 7130-7137.
5. Shannon, R. J.; Taylor, S.; Goddard, A.; Blitz, M. A.; Heard, D. E. Observation of a Large Negative Temperature Dependence for Rate Coefficients of Reactions of OH with Oxygenated Volatile Organic Compounds Studied at 86-112 K. *Phys. Chem. Chem. Phys.* **2010,** *12*, 13511-4.
6. Shannon, R. J.; Caravan, R. L.; Blitz, M. A.; Heard, D. E. A Combined Experimental and Theoretical Study of Reactions between the Hydroxyl Radical and Oxygenated Hydrocarbons relevant to Astrochemical Environments. *Phys. Chem. Chem. Phys.* **2014,** *16*, 3466-3478.
7. Jimenez, E.; Antinolo, M.; Ballesteros, B.; Canosa, A.; Albaladejo, J. First Evidence of the Dramatic Enhancement of the Reactivity of Methyl Formate (HC(O)OCH3) with OH at Temperatures of the Interstellar Medium: A Gas-Phase Kinetic Study between 22 K And 64 K. *Phys. Chem. Chem. Phys.* **2016,** *18*, 2183-2191.
8. Tizniti, M.; Le Picard, S. D.; Lique, F.; Berteloite, C.; Canosa, A.; Alexander, M. H.; Sims, I. R. The Rate of the F + H2 Reaction at Very Low Temperatures. *Nat. Chem.* **2014,** *6*, 141-5.
9. Xu, S.; Lin, M. C. Theoretical Study on the Kinetics for OH Reactions with CH3OH and C2H5OH. *Proc. Combust. Inst.* **2007,** *31*, 159-166.
10. Stark, K.; Werner, H. J. An Accurate Multireference Configuration Interaction Calculation of the Potential Energy Surface for the F+H2→HF+H Reaction. *J. Chem. Phys.* **1996,** *104*, 6515-6530.
11. Frost, M. J.; Sharkey, P.; Smith, I. W. M. Reaction between OH (OD) Radicals and CO at Temperatures down to 80 K: Experiment and Theory. *J. Phys. Chem.* **1993,** *97*, 12254 - 12259.
12. Weston, R. E.; Nguyen, T. L.; Stanton, J. F.; Barker, J. R. HO + CO Reaction Rates and H/D Kinetic Isotope Effects: Master Equation Models with ab initio SCTST Rate Constants. *J. Phys. Chem. A* **2013,** *117*, 821-835.
13. Opansky, B. J.; Leone, S. R. C2H+CH4, CD4 between 150 and 359K. *J. Phys. Chem.* **1996,** *100*, 4888-4892.
14. Dash, M. R.; Rajakumar, B. Abstraction and Addition Kinetics of C2H Radicals with CH4, C2H6, C3H8, C2H4, and C3H6: CVT/SCT/ISPE and Hybrid Meta-DFT Methods. *Phys. Chem. Chem. Phys.* **2015,** *17*, 3142-3156.
15. Agúndez, M.; Wakelam, V. Chemistry of Dark Clouds: Databases, Networks, and Models. *Chem. Rev.* **2013,** *113*, 8710-8737.
16. Ozkan, I.; Dede, Y. Reactions of 1S, 1D, and 3P Carbon Atoms with Water. Oxygen Abstraction and Intermolecular Formaldehyde Generation Mechanisms; An MCSCF Study. *Int. J. Quantum. Chem.* **2012,** *112*, 1165-1184.
17. Hwang, D. Y.; Mebel, A. M.; Wang, B. C. Ab Initio Study of the Addition of Atomic Carbon with Water. *Chem. Phys.* **1999,** *244*, 143-149.
18. Schreiner, P. R.; Reisenauer, H. P. The "Non-Reaction" of Ground-State Triplet Carbon Atoms with Water Revisited. *ChemPhysChem* **2006,** *7*, 880-885.
19. Husain, D.; Kirsch, L. J. Reactions of Atomic Carbon C(23PJ) by Kinetic Absorption Spectroscopy in the Vacuum Ultra-Violet. *J. Chem. Soc. Faraday Trans.* **1971,** *67*, 2025-2035.
20. Husain, D.; Young, A. N. Kinetic Investigation of Ground State Carbon Atoms, C(3P). *J. Chem. Soc. Faraday Trans. 2* **1975,** *71*, 525.
21. Ahmed, S. N.; McKee, M. L.; Shevlin, P. B. An Experimental and Ab Initio Study of the Addition of Atomic Carbon to Water. *J. Am. Chem. Soc.* **1983,** *105*, 3942-3947.
22. CRESU stands for Cinétique de Reaction en Ecoulement Supersonique Uniforme or Reaction Kinetics in a Uniform Supersonic Flow
23. Bergeat, A.; Loison, J.-C. Reaction of Carbon Atoms, C(2p2, 3P) with C2H2, C2H4 and C6H6: Overall Rate Constant and Relative Atomic Hydrogen Production. *Phys. Chem. Chem. Phys.* **2001,** *3*, 2038.
24. Wiese, W. L.; Fuhr, J. R. Accurate Atomic Transition Probabilities for Hydrogen, Helium, and Lithium. *J. Phys. Chem. Ref. Data* **2009,** *38*, 565.
25. Glowacki, D. R.; Liang, C. H.; Morley, C.; Pilling, M. J.; Robertson, S. H. MESMER: An Open-Source Master Equation Solver for Multi-Energy Well Reactions. *J. Phys. Chem. A* **2012,** *116*, 9545-60.
26. Sleiman, C.; Gonzalez, S.; Klippenstein, S. J.; Talbi, D.; El Dib, G.; Canosa, A. Pressure Dependent Low Temperature Kinetics for CN + CH3CN: Competition between Chemical Reaction and Van Der Waals Complex Formation. *Phys. Chem. Chem. Phys.* **2016,** *18*, 15118-15132.
27. Schilke, P.; J., K.; J., L.; G., P. d. F.; Roueff, E. Atomic Carbon in a Dark Cloud: TMC-1. *A&A* **1995,** *294*, L17-L20.

28. Dartois, E. The Ice Survey Opportunity of ISO. *Space Science Rev.* **2005,** *119*, 293-310.
29. Garrod, R. T.; Wakelam, V.; Herbst, E. Non-Thermal Desorption from Interstellar Dust Grains via Exothermic Surface Reactions. *Astron. Astrophys.* **2007,** *467*, 1103-1115.
30. Peters, P. S.; Duflot, D.; Wiesenfeld, L.; Toubin, C. The H + CO -> HCO Reaction Studied by Ab Initio Benchmark Calculations. *J. Chem. Phys.* **2013,** *139*, 164310.
31. Peters, P. S.; Duflot, D.; Faure, A.; Kahane, C.; Ceccarelli, C.; Wiesenfeld, L.; Toubin, C. Theoretical Investigation of the Isomerization of trans-HCOH to H2CO: An Example of a Water-Catalyzed Reaction. *J. Phys. Chem. A* **2011,** *115*, 8983-8989.
32. Shannon, R. J.; Cossou, C.; Loison, J.-C.; Caubet, P.; Balucani, N.; Seakins, P. W.; Wakelam, V.; Hickson, K. M. The Fast C(3P) + CH3OH Reaction as an Efficient Loss Process for Gas-Phase Interstellar Methanol. *RSC Advances* **2014,** *4*, 26342.
33. Hickson, K. M.; Loison, J.-C.; Bourgalais, J.; Capron, M.; Le Picard, S. D.; Goulay, F.; Wakelam, V. The C(3P) + NH3 Reaction in Interstellar Chemistry. II. Low Temperature Rate Constants and Modeling of NH, NH2, and NH3 Abundances in Dense Interstellar Clouds. *Astrophys. J.* **2015,** *812*, 107.
34. Harvey, A. H.; Lemmon, E. W. Correlation for the Vapor Pressure of Heavy Water From the Triple Point to the Critical Point. *J. Phys. Chem. Ref. Data* **2002,** *31*, 173-181.
35. Troe, J. Theory of Thermal Unimolecular Reactions at Low Pressures. I. Solutions of the Master Equation. *J. Chem. Phys.* **1977,** *66*, 4745-4757.
36. Garrett, B. C.; Truhlar, D. G. Semiclassical Tunneling Calculations. *J. Phys. Chem.* **1979,** *83* (22), 2921-2926.
37. Sander, S. P.; J. Abbatt; J. R. Barker; J. B. Burkholder; R. R. Friedl; D. M. Golden; R. E. Huie; C. E. Kolb; M. J. Kurylo; G. K. Moortgat; V. L. Orkin; Wine, P. H. Chemical Kinetics and Photochemical Data for Use in Atmospheric Studies, Evaluation No. 17. *JPL Publication 10-6, Jet Propulsion Laboratory, Pasadena, 2011* http://jpldataeval.jpl.nasa.gov/.
38. Hickson, K. M.; Caubet, P.; Loison, J.-C. Unusual Low-Temperature Reactivity of Water: The CH + H2O Reaction as a Source of Interstellar Formaldehyde? *J. Phys. Chem. Lett.* **2013,** *4*, 2843-2846.
39. Bourgalais, J.; Roussel, V.; Capron, M.; Benidar, A.; Jasper, A. W.; Klippenstein, S. J.; Biennier, L.; Le Picard, S. D. Low Temperature Kinetics of the First Steps of Water Cluster Formation. *Phys. Rev. Lett.* **2016,** *116*, 113401.
40. Hickson, K. M.; Loison, J.-C.; Lique, F.; Kłos, J. An Experimental and Theoretical Investigation of the C(1D) + N2 → C(3P) + N2 Quenching Reaction at Low Temperature. *J. Phys. Chem. A* **2016,** *120*, 2504-2513.
41. Vranckx, S.; Peeters, J.; Carl, S. Kinetics of O(1D) + H2O and O(1D) + H2: Absolute Rate Coefficients and O(3P) Yields between 227 and 453 K. *Phys. Chem. Chem. Phys.* **2010,** *12*, 9213-9221.
42. Hickson, K. M.; Loison, J.-C.; Guo, H.; Suleimanov, Y. V. Ring-Polymer Molecular Dynamics for the Prediction of Low-Temperature Rates: An Investigation of the C(1D) + H2 Reaction. *J. Phys. Chem. Lett.* **2015,** *6*, 4194-4199.
43. Chastaing, D.; Le Picard, S. D.; Sims, I. R.; Smith, I. W. M. Rate coefficients for the Reactions of C(3PJ) Atoms with C2H2, C2H4, CH3CCH and H2C = C = CH2 at Temperatures down to 15 K. *Astron. Astrophys.* **2001,** *365*, 241-247.
44. Zhao, Y.; Truhlar, D. The M06 Suite of Density Functionals for Main Group Thermochemistry, Thermochemical Kinetics, Noncovalent Interactions, Excited States, and Transition Elements: Two New Functionals and Systematic Testing of Four M06-Class Functionals and 12 other Functionals. *Theor. Chem. Acc.* **2008,** *120*, 215-241.
45. Baulch, D. L.; Bowman, C. T.; Cobos, C. J.; Cox, R. A.; Just, T.; Kerr, J. A.; Pilling, M. J.; Stocker, D.; Troe, J.; Tsang, W.; Walker, R. W.; Warnatz, J. Evaluated Kinetic Data for Combustion Modeling: Supplement II. *J. Phys. Chem. Ref. Data* **2005,** *34*, 757-1397.
46. Wagner, A. F. The challenges of Combustion for Chemical Theory. *Proc. Combust. Inst.* **2002,** *29*, 1173-1200.
47. Lee, T. J. Comparison of the T1 and D1 Diagnostics for Electronic Structure Theory: A New Definition for the Open-Shell D1 Diagnostic. *Chem. Phys. Lett.* **2003,** *372*, 362-367.
48. Leininger, M. L.; Nielsen, I. M. B.; Crawford, T. D.; Janssen, C. L. A New Diagnostic for Open-Shell Coupled-Cluster Theory. *Chem. Phys. Lett.* **2000,** *328*, 431-436.
49. Antiñolo, M.; Agúndez, M.; Jiménez, E.; Ballesteros, B.; Canosa, A.; Dib, G. E.; Albaladejo, J.; Cernicharo, J. Reactivity of OH and CH3OH between 22 and 64 K: Modeling the Gas Phase Production of CH3O in Barnard 1b. *Astrophys. J.* **2016,** *823*, 25.

Supplementary Information for

**Quantum Tunneling Enhancement of the C + $H_2O$ And C + $D_2O$ Reactions at Low Temperature**

Kevin M. Hickson, Jean-Christophe Loison, Dianailys Nuñez-Reyes and Raphaël Méreau

correspondence to: kevin.hickson@u-bordeaux.fr

**The SI file includes:**

Experimental Methods

Theoretical Methods

Figs. S1 to S7

Tables S1 to S5

**Experimental Methods**

The CRESU method was used in the present experiments. This technique is well suited to the study of low temperature reactions involving species in the liquid phase at room temperature. As the cold flow is isolated from the reactor walls, condensation issues are avoided and supersaturated concentrations of species with low vapour pressures are attainable. Four Laval nozzles were used in this study allowing different temperatures to be obtained using Ar or $N_2$ as carrier gases and experiments at 296 K were performed by removing the nozzle. All gases were passed directly from cylinders into mass flow controllers to precisely regulate the flows. Individual mass flow controllers were calibrated using the pressure rise at constant volume method. The supersonic flow characteristics (the temperatures, densities and velocities) were calculated from measurements of the impact pressure in the cold flow and the stagnation pressure in the Laval nozzle reservoir. These values are listed in Table S2. Experiments were performed at only a single fixed total density for any given temperature due to the requirement for several different nozzles to produce supersonic flows over a range of densities. $C(^3P)$ atoms were generated by the multiphoton dissociation of $CBr_4$ molecules at 266 nm with approximately 20 mJ of pulse energy. Previous studies [32] under similar conditions have shown that excited state carbon atoms $C(^1D)$ are also produced, at the level of 10-15% with respect to $C(^3P)$. A column of carbon atoms of uniform density was thus created by aligning the photolysis laser along the supersonic flow axis. $CBr_4$ was carried into the reactor using a small carrier gas flow over solid $CBr_4$. The gas-phase concentration of $CBr_4$ was estimated to be lower than $1.1 \times 10^{13}$ cm$^{-3}$ from its saturated vapor pressure. Rate constants for the $C(^3P)$ + $H_2O$ reaction were measured by detecting $C(^3P)$ atoms directly by on-resonance VUV LIF. A pulsed narrow band dye laser operating at 766.5 nm was frequency doubled to produce UV radiation at 383.3 nm which was focused into a cell containing 13.3 kPa of xenon and 48.0 kPa of argon for phase matching mounted on a sidearm of the reactor at the level of the observation axis.

Third harmonic generation of the UV beam allowed us to produce tunable VUV radiation at 127.755 nm which was collimated by a $MgF_2$ lens and allowed to enter the reactor, exciting $C(^3P)$ atoms via the $2s^22p^2\ ^3P_2 \rightarrow 2s^22p3d\ ^3D_3°$ transition within the cold supersonic flow. Atomic fluorescence was focused by a lithium fluoride lens onto the photocathode of the solar blind photomultiplier tube (PMT) and processed by a boxcar integration system. In later product formation measurements, $H(^2S)$ or $D(^2S)$ atoms were detected through VUV LIF at 121.567 nm and 121.534 nm respectively using 28.0 kPa of kryton and 72.0 kPa of argon to generate VUV radiation from a fundamental wavelength around 729 nm in this case.

$H_2O$ was introduced into the gas flow upstream of the Laval nozzle using a controlled evaporation mixing (CEM) system. A 1 litre reservoir maintained at 2 bar relative to atmospheric pressure was connected to a liquid flow meter allowing flows of between 0.1 and 5 g hr$^{-1}$ of liquid water to be passed into an evaporation device heated to 373 K. A small flow of Ar or $N_2$ was also fed into the evaporation system to carry water vapour into the reactor. To determine the gas-phase water concentration, the output of the CEM was passed into a 10 cm absorption cell at room temperature. The attenuation of the 185 nm line of a mercury pen-ray lamp was measured alternatively in the presence and absence of water vapour to yield values of the attenuated and non-attenuated intensities, I and $I_o$ respectively using a channel photomultiplier (CPM) operating in photon counting mode. A narrow band filter centred on 185 nm was placed in front of the CPM to prevent radiation from other mercury lines (notably the intense 254 nm line) from reaching the CPM. The room temperature absorption cross-section of water vapour was taken to be 6.78 × 10$^{-20}$ cm$^2$ [37] so that its concentration could be calculated by the Beer-Lambert law. The output of the cell was connected to the reactor using a heating hose maintained at 353 K to avoid condensation. As the water vapour was diluted by at least a factor of five on entering the nozzle reservoir through mixing with the main carrier gas flow, we assume that no supplementary condensation losses occurred upstream of the Laval nozzle.

$D_2O$ could not be introduced into the experiments in this way, given the need for a large volume of $D_2O$ in the CEM reservoir. Moreover, as the absorption cross-section of $D_2O$ is almost an order of magnitude smaller than the $H_2O$ one at 185 nm, it was no longer possible to determine its concentration spectroscopically. Instead, $D_2O$ vapour was carried into the reactor by passing a small flow of Ar or $N_2$ through room temperature liquid $D_2O$ held in a bubbler at a known pressure. The output of this bubbler was then passed into a cold trap held at 288 K to ensure that the saturated vapour pressure of $D_2O$ at 288 K was attained in the gas flow [34]. The output of the cold trap was then fed into the Laval nozzle reservoir, allowing us to accurately determine $D_2O$ concentrations in the supersonic flow.

**$H_2O$ and $D_2O$ concentrations**

The range of $[H_2O]/[D_2O]$ that could be used in the experiments was severely limited by the formation of clusters in the supersonic flow. Our earlier work on the kinetics of the CH + $H_2O$ reaction [38] allowed us to establish the exploitable range of $[H_2O]/[D_2O]$ in these experiments, listed in Table S2. The kinetics of water cluster formation has recently been investigated at low temperature, using He as the bath gas [39]. To determine whether water cluster formation might have an influence on the present experiments, we performed numerical simulations of our experimental system, including the association rate constants derived in this recent study and all other competing processes. As our carrier gases Ar and $N_2$ are likely to be more efficient than He at promoting cluster formation, we also ran simulations with $H_2O$ association rate constants two times higher than the nominal values. The results obtained suggest that cluster formation is negligible at all temperatures for the $H_2O/D_2O$ concentration ranges used here.

The requirement to use low [H₂O]/[D₂O] values had important consequences for the present measurements. Firstly, as the second-order rate constants for the C + [H₂O]/[D₂O] reactions are small (Table S1), the range of observable pseudo-first-order rate constants (the product of the second-order rate constant and the coreagent concentration) was correspondingly small, potentially leading to large errors in the second-order rate constant determinations. Indeed, the magnitude of the pseudo-first-order diffusional loss rate for C-atoms was comparable to or greater than the supplementary loss induced by reaction with H₂O or D₂O for most experiments, particularly at 106 K. Secondly, the product H-atom intensities were also low, a fact which could have led to large potential errors in the relative product yields. To address both of these issues, we made several improvements to the detection system to increase the signal to noise ratio and improve overall stability. A wavemeter was used to continuously monitor the laser wavelength systematically for all experiments, allowing us to limit the effects of wavelength drift for any series of experiments. The shot to shot energy variability of the probe laser was reduced and the average pulse energy was increased for measurements of both $C(^3P)$ and $H(^2S)$ atoms, increasing the corresponding signal levels. As this energy increase also resulted in an increase in the scattered light background signal detected by the solar blind PMT, the tripling cell was displaced 60 cm further away from the reactor by mounting it on a sidearm containing a system of baffles to reduce the divergence of the VUV probe radiation in the reactor. To avoid supplementary losses of VUV light within the sidearm through absorption by residual gases within the reactor, the sidearm was also continuously flushed with N₂ during the experiments.

**Influence of the $C(^1D)$ + H₂O/D₂O reactions**

It the present experiments, it was clear that the reactions of $C(^1D)$ atoms with both H₂O and D₂O might interfere with both the kinetic and product measurements of the corresponding $C(^3P)$ reactions. As the $C(^1D) + H_2O/D_2O$ reactions are barrierless processes [18], their rate constants could be several orders of magnitude larger than the $C(^3P)$ ones, although no experimental measurements are available in the literature. To improve our overall understanding of the present experiments, we studied the kinetics of the $C(^1D) + H_2O$ reaction at two temperatures, 300 K and 127 K, by following the production of H-atoms. The carrier gas for both of these temperatures was Ar, allowing us to maintain a high concentration of $C(^1D)$ atoms in the flow [40]. At these temperatures H-atom production from the $C(^3P) + H_2O$ tunneling reaction was negligible. The second-order plots obtained in these experiments are shown in Fig. S1.

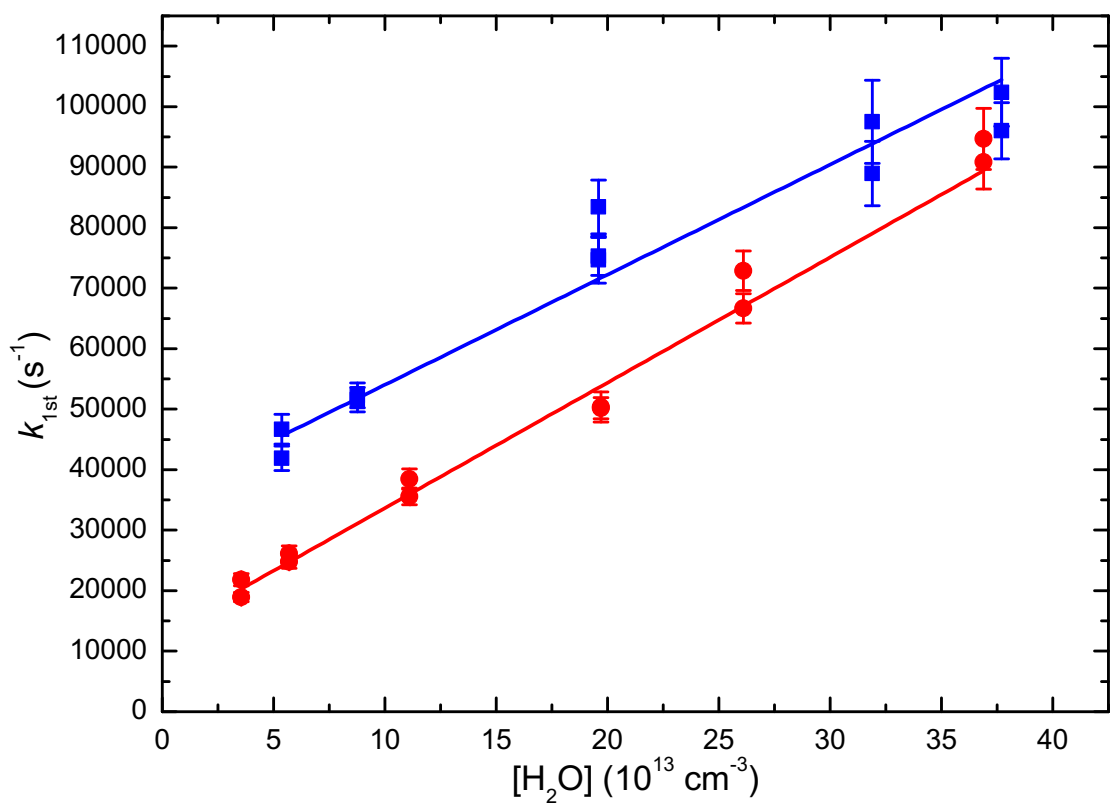

**Figure S1** Measured second-order rate constants for the $C(^1D) + H_2O$ reaction obtained at 296 K (Blue solid squares) and 127 K (Red solid circles). Solid lines represent weighted linear least squares fits to the data, with the second-order rate constants derived from the slopes.

The rate constants are measured to be $(1.8 \pm 0.2) \times 10^{-10}$ cm$^3$ s$^{-1}$ and $(2.1 \pm 0.2) \times 10^{-10}$ cm$^3$ s$^{-1}$ at 296 K and 127 K respectively, with the error bars quoted at the 1σ level with an added 10 % systematic uncertainty. These values show that the rate constant remains high and even increases slightly as the temperature falls. Indeed, a similar dependence is also observed for the O($^1$D) + H$_2$O reaction over the 453 – 227 K range [41].

To check the influence of this reaction on C($^3$P) decays, we performed kinetic measurements for two of our Ar based nozzles (denoted M4Ar and M3Ar in Table S1) with a high concentration of N$_2$ added ([N$_2$] > 9 × 10$^{15}$ cm$^{-3}$) to rapidly relax C($^1$D) atoms. Measurements at 106 K were performed with N$_2$ as the carrier gas. While the nominal temperatures for the M3Ar and M4Ar nozzles are calculated to be (75 ± 2) K and (50 ± 1) K, the use of such large [N$_2$] values shifts the calculated temperatures higher by 2 K. Recent measurements of the quenching of C($^1$D) by N$_2$ down to low temperatures have shown that the rate constants for this process increase to values of around 1.5 × 10$^{-11}$ cm$^3$ s$^{-1}$ at 50 K [40]. The results of such experiments are compared to the results of measurements performed in the absence of N$_2$ in Fig. S2 for the M3Ar nozzle.

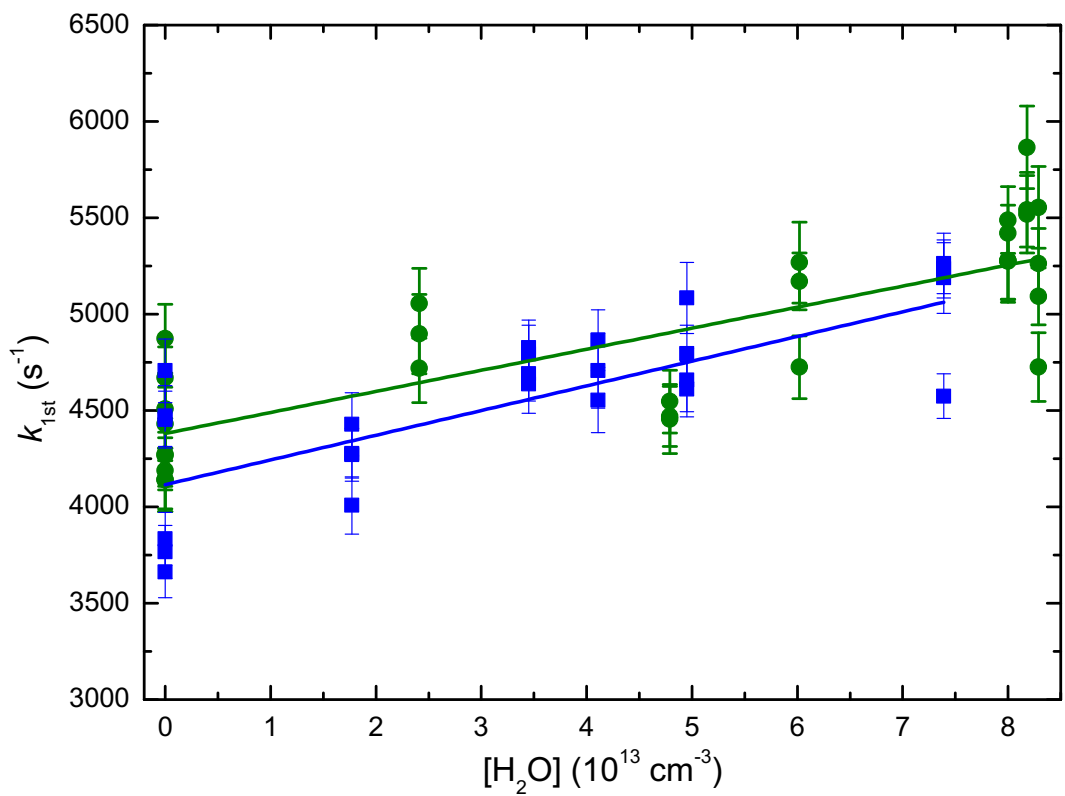

**Figure S2** Measured second-order rate constants for the $C(^3P)$ + $H_2O$ reaction obtained in the presence (blue solid squares, $[N_2]$ = 9 × $10^{15}$ $cm^{-3}$) and absence (green solid circles) of $N_2$ at 77 K and 75 K respectively. Solid lines represent weighted linear least squares fits to the data, with the second-order rate constants derived from the slopes.

It can be seen that the slopes of the two second-order plots are essentially the same, indicating that the $C(^1D)$ + $H_2O$ reaction has little effect on the $C(^3P)$ decay measurements. A similar result was also obtained for experiments performed with the M4Ar nozzle. Although equivalent measurements of the $C(^1D)$ + $D_2O$ reaction were not performed, we expect the rate constants to be similar to the $C(^1D)$ + $H_2O$ ones, given the barrierless nature of these processes. For kinetic experiments performed at 106 K, the carrier gas used was $N_2$ with a flow density of 1.0 × $10^{17}$ $cm^{-3}$. Consequently, $C(^1D)$ atoms were immediately quenched for these measurements. The second-order plots obtained for the C($^3$P) + H$_2$O/D$_2$O reactions at 106 K are displayed in Fig. S3.

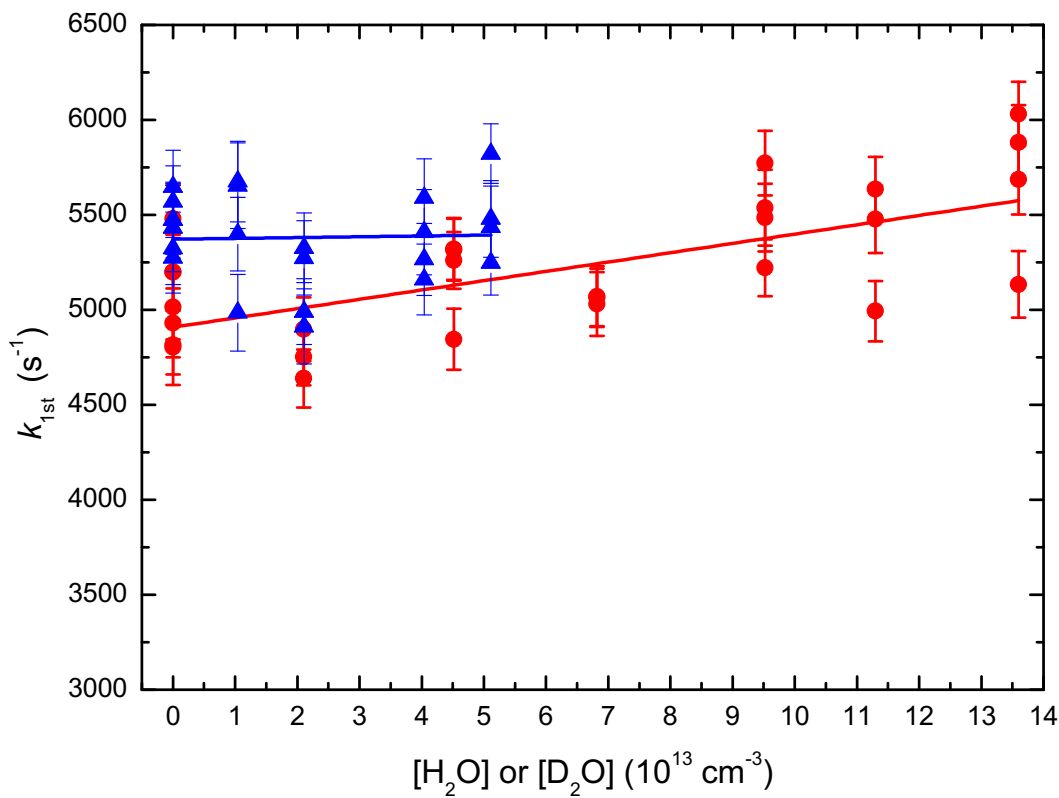

**Figure S3** Measured second-order rate constants for the C($^3$P) + H$_2$O reaction (red solid circles) and the C($^3$P) + D$_2$O reaction (blue solid triangles) at 106 K. Solid lines represent weighted linear least squares fits to the data, with the second-order rate constants derived from the slopes.

The derived second-order rate constant for the C($^3$P) + D$_2$O reaction is very small indeed, with the value of 4 × 10$^{-13}$ cm$^3$ s$^{-1}$ from a non-linear least squares fit to the datapoints. Given the small range of D$_2$O concentrations used and the relatively large scatter of the measured pseudo-first-order rates, we prefer to give an upper limiting value for the rate constant of 3 × 10$^{-12}$ cm$^3$ s$^{-1}$ at 106 K, which corresponds to the sum of the upper bound of the statistical error and the nominal rate constant. For the C($^3$P) + H$_2$O reaction at 106 K, a non-linear least squares fit to the data in Fig. S3 yields a second-order rate constant of $(4.9 \pm 1.1) \times 10^{-12}$ cm$^3$ s$^{-1}$ where the quoted error bars represent the statistical 1σ uncertainty alone. To reflect the high scatter in the measured pseudo-first-order rates we estimate an additional systematic uncertainty of 50 % of the nominal rate, yielding a final value of $(4.9 \pm 2.7) \times 10^{-12}$ cm$^3$ s$^{-1}$. For all of the other lower temperature kinetic measurements, we apply a systematic uncertainty of 20 % on the measured rate constant values.

**Relative product yields of the C($^3$P) + H$_2$O/D$_2$O reactions**

As the C($^1$D) + H$_2$O/D$_2$O reactions are much faster than the corresponding C($^3$P) ones, quantitative measurements of H/D-atom production by C($^3$P) reactions were only possible when N$_2$ was added to the flows. N$_2$ concentrations around $2.1 \times 10^{16}$ cm$^{-3}$ were used to rapidly quench C($^1$D) atoms at 52 K. The pseudo-first-order rate constant for C($^1$D) loss under these conditions is equal to $3.2 \times 10^5$ s$^{-1}$, indicating that C($^1$D) atoms are entirely removed from the flow within the first 10 microseconds. A typical H-atom formation curve is shown in Fig. S4 for experiments performed at 52 K, alongside the corresponding reference reaction (C($^3$P) + C$_2$H$_4$) one with a known H-atom yield of 0.92 ± 0.04 at 300 K [23].

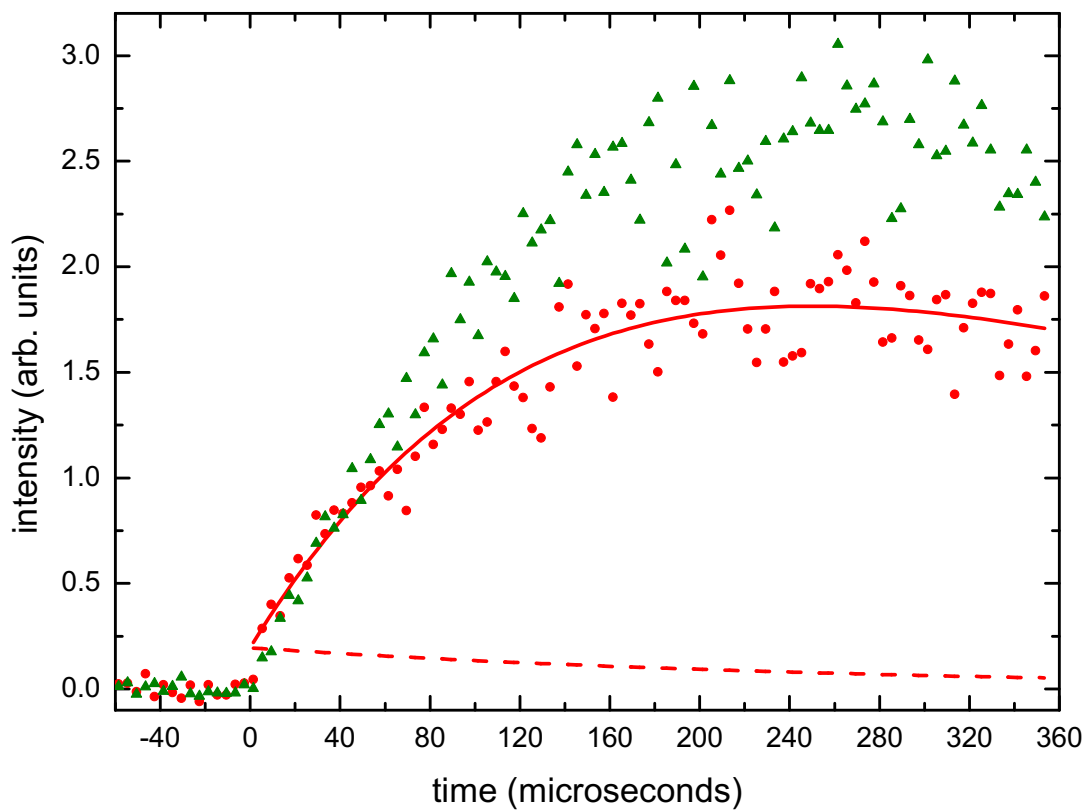

**Figure S4** H-atom formation curves for the target $C(^3P) + H_2O$ (red solid circles) and reference $C(^3P) + C_2H_4$ reaction (green solid triangles) recorded at 52 K with $[H_2O] = 1.8 \times 10^{13}$ cm$^{-3}$ and $[C_2H_4] = 1.5 \times 10^{12}$ cm$^{-3}$. The solid red line represents the biexponential fit used to determine the instantaneous H-atom signal from the $C(^1D) + H_2O$ reaction. The dashed red line represents the temporal dependence of the instantaneous H-atom signal, using a H-atom diffusional loss rate taken from recent work under similar conditions.

It can be seen that despite our use of high $[N_2]$ values, as the rate constant of the $C(^1D) + H_2O$ reaction is almost an order of magnitude larger than the $C(^3P) + H_2O$ one, there remains a small instantaneous production of H-atoms from the $C(^1D) + H_2O$ reaction during the first few microseconds, corresponding to 10-15 % of the peak H-atom signal. To obtain more precise values of the relative product yields, it was therefore necessary to subtract this signal. To estimate the initial H-atom contribution from the C($^1$D) + $H_2O$ reaction, we performed a biexponential fit to the data [42] including a non-zero contribution to the offset. This fit is shown by the solid red line in Fig. S4. These H-atoms are slowly lost by diffusion as shown by the dashed red line in Fig. S4, with the first-order loss rate taken from recent H-atom formation experiments performed under similar conditions [42]. The final H-atom formation curve, an example of which is shown in Fig. 4 of the main article, is obtained by subtracting the H-atom diffusional loss profile from the red datapoints in Fig. S4. No correction was required for H-atom yields from the the C($^3$P) + $C_2H_4$ reaction ($k_{C+C2H4}$(50 K) = 3.6 × $10^{-10}$ $cm^3$ $s^{-1}$ [43]) as the corresponding C($^1$D) + $C_2H_4$ reaction is expected to occur at a similar rate, so its H-atom contribution is likely to be negligibly small compared to the C($^3$P) + $C_2H_4$ one. Indeed, no instantaneous H-atom production was ever observed when $C_2H_4$ was added. To obtain quantitative product yields, [$H_2O$] and [$C_2H_4$] were fixed to obtain the same pseudo-first-order production rates for the target and reference reactions (considering our measured rate constants for the C($^3$P) + $H_2O$ reaction and the literature values for the C($^3$P) + $C_2H_4$ reaction [43]). Several pairs of H-atom formation curves were then recorded and the individual curves were simply divided to obtain the signal ratio as a function of time as shown in Fig. S5.

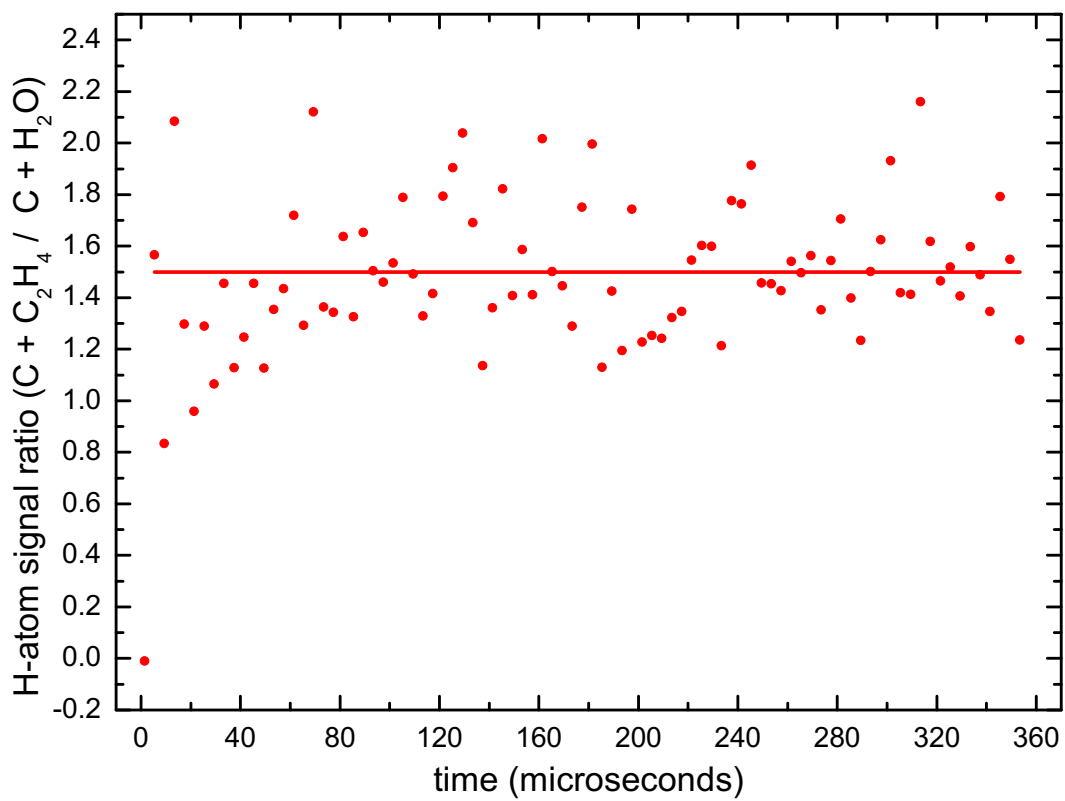

**Figure S5** Ratio of the H-atom signals from the target C($^3$P) + $H_2O$ and reference C($^3$P) + $C_2H_4$ reactions as a function of time. The red solid line represents the mean product yield averaged over all time points.

The relative product yields were then obtained by performing a linear fit (with the slope set to zero) to the individual datapoints represented by a solid red line in Fig. S5. As an added precaution, the entire experiment was also repeated for different higher values of the pseudo-first-order production rate by increasing the $H_2O$ and $C_2H_4$ concentrations. An equivalent procedure was used to compare D-atom yields from the C($^3$P) + $D_2O$ reaction to the corresponding H-atom yield of the C($^3$P) + $C_2H_4$ reference reaction (see Figs. S6 and S7).

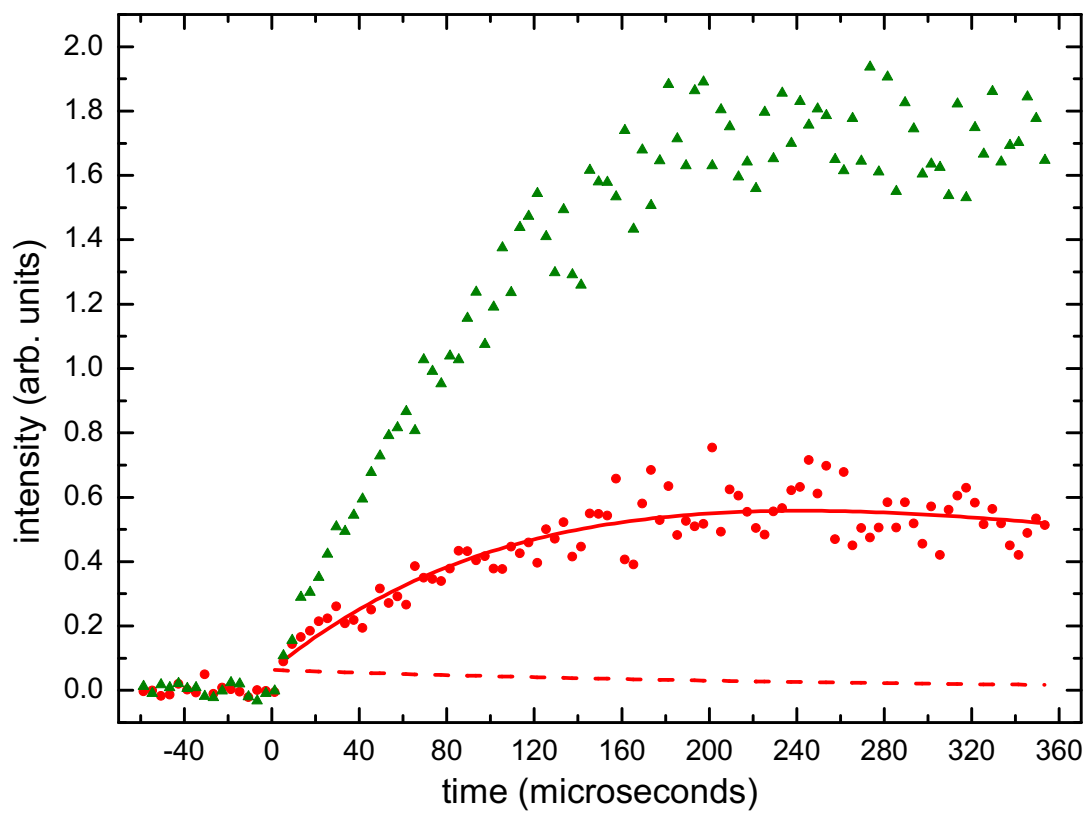

**Figure S6** H-/D-atom formation curves for the target C($^3$P) + D$_2$O (red solid circles) and reference C($^3$P) + C$_2$H$_4$ reaction (green solid triangles) recorded at 52 K with [D$_2$O] = 2.0 × 10$^{13}$ cm$^{-3}$ and [C$_2$H$_4$] = 1.5 × 10$^{12}$ cm$^{-3}$. The solid red line represents the biexponential fit used to determine the instantaneous D-atom signal from the C($^1$D) + D$_2$O reaction. The dashed red line represents the temporal dependence of the instantaneous D-atom signal, assuming the same diffusional loss as the one for H-atoms measured in recent work under similar conditions.

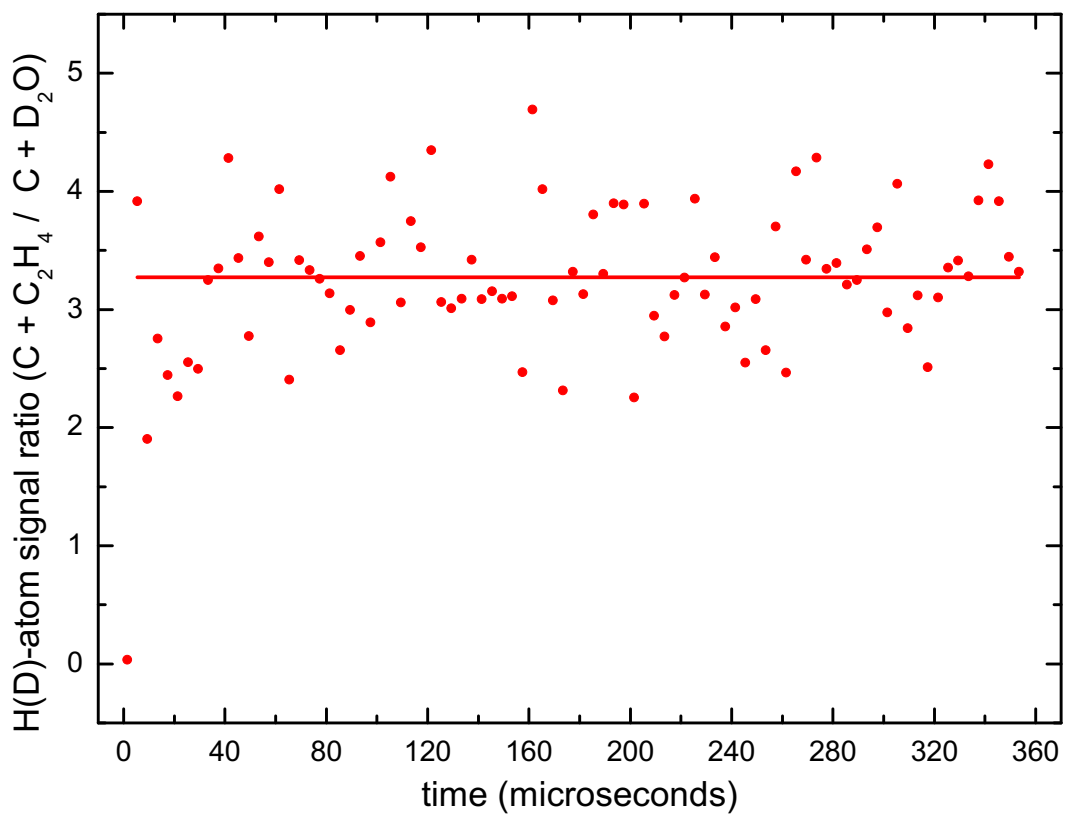

**Figure S7** Ratio of the H-/D-atom signals from the target $C(^3P) + D_2O$ and reference $C(^3P) + C_2H_4$ reactions as a function of time. The red solid line represents the mean product yield averaged over all time points.

In this case, as it was also necessary to move between the D-atom and H-atom transitions at 121.534 nm and 121.567 nm (using the wavemeter to precisely fix the appropriate wavelength), the probe laser pulse energy was set to the same value for each product pair measurement. As the line strength factors are the same for D- and H-atoms [24], no other corrections were required to the measured intensities. Nevertheless, as an additional verification, experiments looking at the H-/D- atom yields of the $C(^1D) + H_2/D_2$ reactions resulted in ratios of unity for fixed probe laser pulse energies. All the ratios determined from individual pairs of decays are given in Table S3 for both reactions after multiplication by 0.92 to account for the measured yield of the reference reaction.

H-/D-atom product yields were only derived from data recorded at low $H_2O$ or $D_2O$ concentrations (< $3.8 \times 10^{13}$ $cm^{-3}$) for three main reasons. Firstly, such low values allowed us to neglect the potential absorption of the VUV excitation laser and VUV fluorescence emission by residual gas in the reactor. Secondly, it allowed us to minimize the contributions of the $C(^1D)$ + $H_2O/D_2O$ reactions to the H-/D-atom formation curves. Thirdly, any small contributions to the H-/D-atom signals from $H_2O/D_2O$ clusters that might be present in the flow (which may or may not be able to react rapidly with ground state atomic carbon) were entirely negligible. The average H- and D-atom product yields given in the main article were derived from 11 and 9 individual pairs of formation curves respectively.

While efforts were made to record relative product yields at higher temperatures, these measurements were hampered by several issues. At 106 K in particular, comparison of the extremely weak H-/D-atom formation curves with H-atom yields from the reference $C(^3P)$ + $C_2H_4$ reaction led to large product yield variations. As these measurements rely on the use of identical pseudo-first-order production rates for the target and reference reactions, given the large errors on the measured second-order rate constants, such variations were not unexpected. Indeed, further attempts to overcome these problems by using much larger concentrations of $H_2O$ and $D_2O$ also failed with these measurements potentially being affected by cluster formation in the supersonic flow. At 77 K, the instantaneous production of H-/D-atoms from the $C(^1D)$ + $H_2O/D_2O$ was considerably larger (the $C(^1D)$ + $N_2$ quenching rate constant decreases at higher temperature), so it became difficult to extract reliable product yields under these conditions.

**Theoretical Methods**

**Quantum chemical calculations**

There have been several previous calculations of the reaction of carbon atoms with water [16-18] the most complete one being the study of Schreiner & Reisenauer [18]. To perform the statistical calculations presented here, we performed new ab-initio calculations at a higher level than these earlier studies, namely at the CCSD(T)/aug-cc-pVQZ level, to describe all stationary points on the triplet surfaces. To estimate the potential uncertainties of the calculations we also performed MP2 calculations at the aug-cc-pVQZ level, DFT calculations (M06-2X level with the cc-pVQZ basis set; this highly nonlocal M06-2X functional developed by Truhlar *et al.*[44] is well suited for structures and energetics of the transition states), CASPT2 and Davidson corrected multi-reference configuration interaction (MRCI + Q) with complete active space self-consistent field (CASSCF) wavefunctions, to take into account the eventual multi-configurational aspect of this reaction.

CASPT2, MRCI+Q and CCSD(T) calculations were carried out with the MOLPRO 2010 package whereas the MP2 and DFT calculations were performed with Gaussian09. The CASSCF and MRCI calculations were performed at full valence, namely with 8 electrons distributed in 10 orbitals with the 1s orbitals of carbon and oxygen kept doubly occupied but fully optimized. The CASSCF calculations lead to mono-configurational wavefunctions for the adduct and TS1 transition states. Then CCSD(T) calculations should be reliable as indicated by the T1 and D1 diagnostic values (T1 values are in the 0.004-0.020 range and D1 values are in the 0.010-0.044 range). The geometries for all stationary points were fully optimized at the aug-cc-pVQZ level for each method and frequencies were calculated for each method using a smaller basis set (aug-cc-pVTZ) at the CCSD(T) and MRCI+Q levels. For each TS, characterized by one imaginary frequency (first-order saddle points) on the PES, we determined the minimum energy pathways (MEPs) performing intrinsic reaction coordinate analyses (IRC) at the MP2 and DFT levels. A schematic diagram of the PESs is shown in Figure 1 of the main article.

The calculated stationary point energies are summarized in Table S4 for the prereactive complex and TS1 at various levels. In Table S5, we list all the stationary points of the triplet PESs at the CCSD(T)/aug-cc-pVQZ level as well as the geometries, frequencies and structures of the various species.

The C($^3P_{0,1,2}$) + H$_2$O(X$^1A_1$) reaction leads to 3 surfaces in the entrance valley, one with $^3A'$ symmetry and two with $^3A''$ symmetry when the C-atom approaches in C$_s$ geometry or 3 $^3A'$ surfaces in C$_1$ geometry (no symmetry). At the MRCI+Q level, with the CASSCF geometry optimized for non-relaxed H$_2$O (the geometry of the isolated molecule), only one of the three surfaces is attractive leading to a van der Waals complex. This complex is strongly bound (between -17.9 and -48.9 kJ mol$^{-1}$ depending on the calculation level), with the C-atom attached to the O-atom of water with a C-O distance between 1.76 Å and 2.45 Å depending on the method used. Further evolution of the complex leads to the $^3$HCOH intermediate through TS1 corresponding to insertion of the C-atom into the O-H bond, or directly to H + HOC through TS4. We were unable to locate the TS for the pathway leading to $^3$CO + H$_2$. However, as these products are almost iso-energetic with the C + H$_2$O reagents, if this TS exists, is likely to be at very high energy and should not play any role in the present low temperature measurements. As a result, this pathway is not considered further in these calculations.

Our results are in good agreement with previous calculations[16-18] as well as with thermochemical data when they exist (-213.8 kJ mol$^{-1}$ [45] compared to -206.4 kJ mol$^{-1}$ (this study) for the HCO + H exit channel). However, as shown in Table S4, there are significant variations in the calculated energies of the complex and TS1 depending on the method used. It should be noted that the C-O distance calculated at the CASSCF level (2.45 Å) does not correspond to the minimum energy configuration for MRCI+Q calculations so that the MRCI+Q results may not correspond exactly to the stationary point energies despite various adjustments of the C-O length in the complex. The precision of the various methods is coherent with expectations on their accuracies. Indeed, as shown by Wagner [46], DFT usually underestimates barrier energies whereas MRCI usually overestimates them. CASPT2 calculations are likely to underestimate barrier energies, as do MCQDPT2 calculations [16]. CCSD(T) should lead to the most accurate values considering the T1 and D1 diagnostic values [47-48] in agreement with the fact that CASSCF calculations lead to mono-configurational wavefunctions. As a result, the expected uncertainty on the complex and TS energies is difficult to estimate but it could be greater than 10 kJ mol$^{-1}$.

**Master Equation Calculations**

Rate constants and product branching ratios were calculated using the Master Equation solver MESMER[25], an open source program which uses matrix techniques to solve the energy grained chemical master equation for a series of unimolecular systems connected by various transitions states and local minima leading from reagents to products. Briefly, the phase space for all wells on the PES was divided into grains of a set size and isoenergetic, microcanonical rate coefficients for reactive processes between different species were calculated using RRKM theory, while energy transfer probabilities between different grains of the same species were obtained using an exponential down model [35]. For all species, external rotations were considered to be rigid rotors whilst vibrational modes were treated as harmonic oscillators.

For the current calculations, a grain size of 20 cm$^{-1}$ was used above 150 K although this was reduced to values as low as 10 cm$^{-1}$ at the lowest temperatures. Harmonic frequencies and rotational constants were taken from the CCSD(T) calculations but given the large range of energies for the stationary points between the different levels of theory used here, the energy of the complex was varied during the calculations as well as the width of the barrier connecting the complex and HCOH

(the IRC was calculated at the MP2 level). During the procedure to reproduce the experimental data, the same adjustments were applied to both the C + $H_2O$ and C + $D_2O$ systems for the energies of the prereactive complexes as well as to the width of the barrier to maintain a consistent description of these two reactions.

The barrierless reaction step describing complex formation was treated using the inverse Laplace transform (ILT) method to convert the high pressure limiting canonical rate coefficients for the initial association in Arrhenius form into microcanonical rate coefficients. Considering the fine structure of ground state atomic carbon, only the $^3P_0$ state and two levels of the $^3P_1$ states of the C-atom lead to reaction as mentioned above. Using capture theory, this leads to a rate constant $k(T)$ = $7.0e^{-10} \times (T/298)^{0.17} \times (1 + 2 \times \exp(-23.6/T))/(1 + 3 \times \exp(-23.6/T) + 5 \times \exp(-62.40/T))$ for the initial complex formation step dominated by dispersion interactions.

The measured product branching ratios allowed us to provide certain additional constraints for the initial parameters of the MESMER statistical calculations. As the H-atom yield is less than unity for both the C + $H_2O$ and C + $D_2O$ reactions, some complex stabilization clearly occurs. To treat collisional energy transfer, the average downward energy transfer value, $\Delta E_{down}$, was taken to be equal to $400*(T/300)^{0.85}$ $cm^{-1}$ in the initial trial model although this was found to be incompatible with the experimental results (see below for details). The Lennard-Jones parameters σ and ε were set respectively to 3.47 Å and 114 $cm^{-1}$ for Ar, to 3.90 Å and 82 $cm^{-1}$ for $N_2$ and to 5.0 Å and 150 $cm^{-1}$ for the C…$H_2O$/$D_2O$ prereactive complexes. The pressures used in the calculations were set to be equal to the experimental values (see Table S2).

When using the CCSD(T) well depth for the complex, the WKB method for tunneling using the IRC calculated at the MP2 level (adjusted to take into account the ZPE variation) and the standard energy transfer value, the master equation calculations predict that some complex stabilization occurs, but bimolecular product formation is negligible. In order to increase the yield of the reactive pathway leading to bimolecular product formation (H + HCO), it was necessary to either substantially decrease the TS1 barrier height or decrease its width. Decreasing the barrier height leads to a rate constant which presents little temperature dependence, with a significantly greater value than the upper limit measured by Husain and Young [20]. Consequently, we maintained this barrier height at the calculated value (+32.9 kJ mol$^{-1}$ for $H_2O$ and +35.1 kJ mol$^{-1}$ for $D_2O$) and varied the barrier width instead. To reproduce the experimental results, it was necessary to reduce the width by a factor close to 2. Moreover, it was also necessary to increase either the prereactive complex well depth and/or the collisional energy transfer efficiency. Both of these adjustments lead to longer complex lifetimes, thereby enhancing the tunneling process at low temperature as well as increasing the probability for complex stabilization through three body collisions. In the absence of either of these adjustments, the major fate of the prereactive complex is dissociation back to reagents, even at low temperature. This is almost certainly due to the fact that the C + $H_2O$ system is relatively small with a low density of rovibrational levels. Increasing only the complex well depth, to -60.0 kJ mol$^{-1}$ (instead of -28.9 kJ/mol at the CCSD(T) level), allows reasonable agreement to be obtained between the calculations and the measured reaction rates/branching ratios. Good agreement could also be obtained by adjusting the well depth to -46 kJ mol$^{-1}$ and by using a temperature independent $\Delta E_{down}$ value of 400 cm$^{-1}$. These values were the ones used to obtain the calculated rates presented in Figure 3 of the main article. As a comparison, in the recent study by Sleiman et al. [26] of the CN + $CH_3CN$ reaction, a temperature independent value of 800 cm$^{-1}$ was used for the $\Delta E_{down}$ parameter.

Using these parameters, the branching ratio between complex stabilization and H-atom product formation is not completely reproduced for the experimental conditions at 52 K for the C + $H_2O$ reaction (92% for the calculated product formation instead of the measured value of 58%) while it is well reproduced for the C + D$_2$O reaction (30% for the calculated product formation instead of the measured value of 29%). The calculated values were predicted to decrease with increasing pressure, due to an increase of the total rate constant arising from more efficient complex stabilization (with only a weak corresponding effect on the bimolecular product formation rate). As mentioned in the main article, we do not consider the possible contributions of HCO or DCO fragmentation in the calculated H-/D-atom branching ratios. As the observed H-/D-atoms are clearly identified as products of the C + water reaction (and not from competing secondary reactions) with well-defined yields, this strongly suggests that statistical theory can only provide semi-quantitative explanations for the experimental observations in this instance.

It is interesting to compare our calculations with other recent studies of OH radical reactions with various oxygenated volatile organic compounds [3-4, 6, 49] as well as the CN + CH$_3$CN reaction [26] at low temperature. All these processes are characterized by prereactive complex formation associated with a notable barrier on the reactive pathways. Moreover, all these reactions show similar behavior with a strong increase of the rate constant at low temperature. However, the actual mechanisms are considered to be different in many of these systems. In some cases, (notably for the OH + CH$_3$OH reaction [3]) the main process was thought to be tunneling through the barrier leading to reaction. In other cases (the OH + acetone reaction [6], the OH + dimethylether reaction [6], and OH + ethanol and propan-2-ol [4]) the experimental conditions were considered to favor complex stabilization rather than bimolecular product formation. It should be noted that in many of these previous studies, the RRKM calculations were generally unable to reproduce the experimental results without large adjustments. For their study of the OH + acetone reaction, Shannon et al. [6] increased the well depth for the prereactive complex by 50% to reproduce the experimental data.

In their investigation of the OH + methanol reaction [3], one imaginary frequency was divided by 4.6 leading only to qualitative agreement between the calculated and experimental rate constant values. In a recent study of the CN + $CH_3CN$ reaction, Sleiman et al. [26] used a large value for the $\Delta E_{down}$ parameter to stabilize efficiently the prereactive complex. Nevertheless, they were unable to reproduce the rate constant temperature dependence below 100 K. In the context of this previous work, where large deviations from the initial ab-initio values and standard parameters were required, the modifications applied in the present work appear to be quite reasonable.

The $C(^3P)$ + $H_2O/D_2O$ reactions display large negative temperature dependences which can be reproduced theoretically by increasing the prereactive complex lifetime as described above. As the experimental H-/D-atom yields demonstrate that complex stabilization is non-negligible, the statistical calculations have been used to extrapolate the rate constant of the $C(^3P)$ + $H_2O$ reaction to temperatures and pressures more representative of dense interstellar clouds although it was not possible to extend these calculations below 30 K due to numerical difficulties. While the statistical calculations predict that complex stabilization is a pressure dependent process, the bimolecular product channel is found to depend only slightly on the total pressure. Indeed, three body collisions increase the complex lifetime, thereby increasing the possibility of tunneling through the barrier. Here the experimental total rate constant at 52 K is multiplied by the H-atom branching ratio of 0.58 to yield the reactive (tunneling) rate alone. The calculations predict that the total rate constants measured at 77 and 106 K are essentially the same as the tunneling rate already (complex association is negligible at these temperatures). We also use the experimental upper limit of Husain and Young at 300 K [20] of $1 \times 10^{-12}$ $cm^3$ $s^{-1}$ to further constrain the fit. These data are fitted using an expression of the type $a \times (T/300)^b$ leading to a = $1.26 \times 10^{-12}$ $cm^3$ $s^{-1}$ and b = -1.59 in the 52 – 300 K range. Then, as the RRKM calculations show that the tunneling rate constant has a small pressure dependence, we multiply the fitted rate constant expression by the ratio between the calculated zero pressure value and the calculated value at the pressure used in the 52 K experiment. This ratio is equal to 0.74 leading to a modified formula with a = $9.3 \times 10^{-13}$ and b= -1.59. At 10 K, this expression yields a rate constant of $2 \times 10^{-10}$ cm$^3$ s$^{-1}$. Given the predicted high abundance of atomic carbon in dense interstellar clouds and the large rate constant, this process should become the most efficient mechanism for H$_2$O loss in current models.

**Table S1** Measured rate constants for the $C(^3P) + H_2O$ and $C(^3P) + D_2O$ reactions

| $T$ (K) | N* | $[H_2O]$† | $k_{C+H2O}$‡,§ | $T$ (K) | N* | $[D_2O]$† | $k_{C+D2O}$‡,§ |
|---|---|---|---|---|---|---|---|
| 50 | 22 | 0-8.5 | 34.7±7.4 | 52 | 26 | 0-8.6 | 30.1±7.8 |
| 52 | 28 | 0-8.3 | 37.0±7.8 | 52 | 26 | 0-8.4 | 30.4±7.4 |
| 75 | 30 | 0-8.3 | 10.9±2.7 | 77 | 50 | 0-8.5 | 9.7±2.7 |
| 77 | 25 | 0-7.4 | 12.8±3.4 | 106 | 23 | 0-5.1 | < 3.0 |
| 106 | 28 | 0-13.6 | 4.9±2.7 | | | | |

*Number of individual measurements. †Concentration range ($10^{13}$ cm$^{-3}$). ‡Second–order rate constant ($10^{-12}$ cm$^3$ s$^{-1}$). §Errors are cited at the 1σ level with an added systematic uncertainty (see text for details).

**Table S2** Continuous supersonic flow characteristics

| Laval nozzle | Mach2 N$_2$ (using Ar) | Mach3 N$_2$ | Mach3 Ar | Mach3 Ar | Mach4 Ar | Mach4 Ar |
|---|---|---|---|---|---|---|
| Mach number | 2.0 ± 0.03* | 3.0 ± 0.02 | 3.0 ± 0.1 | 3.0 ± 0.1 | 3.9 ± 0.1 | 3.9 ± 0.1 |
| Carrier gas | Ar | N$_2$ | Ar (7% N$_2$) | Ar | Ar (7% N$_2$) | Ar |
| Density ($10^{16}$ cm$^{-3}$) | 12.6 | 10.3 | 14.7 | 14.7 | 25.9 | 25.9 |
| Temperature (K) | 127 ± 2$^a$ | 106 ± 1 | 77 ± 2 | 75 ± 2 | 52 ± 1 | 50 ± 1 |
| Mean flow velocity (ms$^{-1}$) | 419 ± 3$^a$ | 626 ± 2 | 479 ± 3 | 479 ± 3 | 505 ± 1 | 505 ± 1 |

*The errors on the Mach number, temperature and mean flow velocity, cited at the level of one standard deviation from the mean are calculated from separate measurements of the impact pressure using a Pitot tube as a function of distance from the Laval nozzle and the stagnation pressure within the reservoir.

**Table S3** Individual product yields for the C($^3$P) + H$_2$O and C($^3$P) + D$_2$O reactions at 52 K

| [H$_2$O] (cm$^{-3}$) | H-atom yield C + H$_2$O / H-atom yield C + C$_2$H$_4$ | [D$_2$O] (cm$^{-3}$) | D-atom yield C + D$_2$O / H-atom yield C + C$_2$H$_4$ |
|---|---|---|---|
| 3.3 × 10$^{13}$ | 0.65±0.01* | 3.8 × 10$^{13}$ | 0.31±0.01* |
| | 0.54±0.01 | | 0.29±0.01 |
| | 0.55±0.01 | | 0.32±0.01 |
| | 0.56±0.01 | | 0.29±0.01 |
| | 0.52±0.01 | | 0.25±0.01 |
| 1.8 × 10$^{13}$ | 0.55±0.01 | 2.0 × 10$^{13}$ | 0.32±0.01 |
| | 0.58±0.01 | | 0.25±0.01 |
| | 0.60±0.01 | | 0.29±0.01 |
| | 0.60±0.01 | | 0.28±0.01 |
| | 0.61±0.01 | | |
| | 0.62±0.01 | | |
| | **0.58±0.04†** | | **0.29±0.02†** |

*Individual product yields obtained from fits to the temporal H-/D- atom yields such as those shown in Figs. S5 and S7. The uncertainties are quoted at the 1σ level. †Mean product yields with uncertainties quoted at the 95% confidence level including the errors associated with the reference reaction.

**Table S4** Calculated energies of the H$_2$O…C prereactive complex and TS1 by different methods (at 0 K including ZPE with respect to the C + H$_2$O reagent asymptote).

| | Complex | TS1 |
|---|---|---|
| MRCI+Q/av5z//CASCCF/av5z | -17.9 | +33.6 |
| CASPT2/av5z//CASCCF/av5z | -48.9 | +0.2 |
| MP2/vqz | -33.2 | +29.8 |
| CCSDT/avqz | -28.9 | +32.9 |
| M06/vqz | -48.4 | +6.8 |

**Table S5** Energies (in kJ/mol at 0 K including ZPE with respect to the C + H$_2$O energy at the CCSD(T)/aug-cc-pVQZ level), geometries and frequencies (in cm$^{-1}$, unscaled, calculated at the CCSD(T)/aug-cc-pVTZ level) of the various stationary points of the C + H$_2$O reaction on the triplet surface as well as the stationary points of the C + D$_2$O reaction used in the RRKM calculations.

| Species | Energies (kJ/mol) | Geometries | Frequencies (cm$^{-1}$) |
|---|---|---|---|
| H$_2$O | 0 | O  0.00000000  0.00000000  -0.06578856<br>H  0.00000000  0.75754244  0.522142971<br>H  0.00000000  -0.75754244  0.5221429713 | 1645, 3810, 3920 |
| D$_2$O | 0 | identical to H$_2$O | 1204, 2747, 2872 |
| 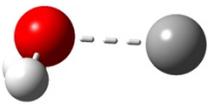<br>H$_2$O…C | -28.9 | O  -0.06649886  0.00000014  0.73931770<br>H  0.47330244  0.76821292  0.95626321<br>H  0.47329692  -0.76822168  0.95624485<br>C  0.00914383  0.00000054  -1.14531122 | 293, 534, 546, 1628, 3755, 3862 |
| D$_2$O…C | -30.3 | identical to H$_2$O…C | 290, 397, 401, 1191, 2706, 2832 |
| 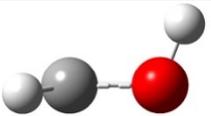<br>HCOH | -246.7 | O  -0.06978105  0.02086028  -0.58617497<br>H  0.65021392  -0.47361658  -0.99449964<br>C  0.08005466  0.05848200  0.74389129<br>H  -0.49651582  -0.55440039  1.43457508 | 413, 1092, 1179, 1287, 3099, 3740 |
| DCOD | -246.5 | identical to HCOH | 307, 839, 896, 1247, 2282, 2722 |
| 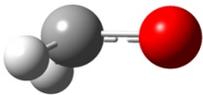<br>³H₂CO | -265.8 | C  -0.06516505  -0.00000003  -0.66646946<br>O   0.01274256  -0.00000001   0.64238782<br>H   0.28713206   0.92748342  -1.12747500<br>H   0.28713230  -0.92748274  -1.12747566 | 754, 878, 1250, 1325, 2988, 3096 |
| 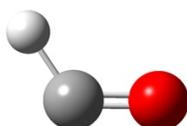<br>HCO (+ H) | -206.4 | C   0.00000000  -0.10157871  -0.62026151<br>O   0.00000000   0.02935911   0.55090499<br>H   0.00000000   0.74442292  -1.35344205 | 1108, 1881, 2702 |
| 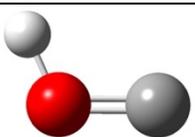<br>HOC (+ H) | -32.8 | O   0.00000000  -0.06816339  -0.51254827<br>C   0.00000000   0.02344327   0.76301534<br>H   0.00000000   0.80262343  -0.95651768 | 1143, 1380, 3478 |
| CO (+ H + H) | -147.6 | C   0.00000000   0.00000000  -0.64649832<br>O   0.00000000   0.00000000   0.48533641 | 2143 |
| 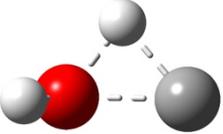 TS1 (H) (H₂O…C →HCOH) | +32.9 | O  -0.02022141   0.06292576  -0.67147866<br>H   0.62086765  -0.60919265  -0.94972510<br>C   0.03745507   0.01152888   0.96765368<br>H  -0.74621493  -0.52703362   0.07741852 | 1514i,<br>641, 851, 1190<br>2185, 3719 |
| TS1 (D) | +35.1 | identical to TS1 (H) | 1142i,<br>613, 629, 867,<br>1601, 2707 |
| 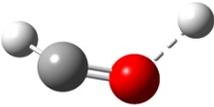 TS2 (HCOH→HCO + H) | -134.6 | C  -0.09799325   0.04698585  -0.68487850<br>H   0.73855721   0.10530504  -1.40605090<br>O   0.02577260  -0.10557536   0.51572022<br>H   0.02007009   1.01062978   1.38111055 | 2145i,<br>641, 762, 1115<br>1652, 2875 |
| 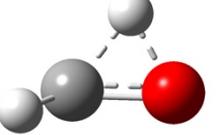 TS3 (HCOH →³H₂CO) | -118.0 | C  -0.08951459  -0.02159168  -0.71302878<br>H   0.61316804   0.43378083  -1.41541008<br>O   0.00909786   0.04923353   0.62085489<br>H   0.30910850  -0.95798786   0.05707807 | 1990i,<br>782, 1061, 1276,<br>2449, 3056 |
| 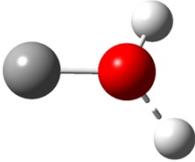<br>TS4<br>(C+$H_2O$→HOC+H) | +93.5 | O   0.03009709  -0.07518982  -0.51246764<br>H   0.74397174   0.50065478  -0.85468233<br>C  -0.02693450   0.02761462   0.86219231<br>H  -0.90075213   0.36379421  -1.28494595 | 2871i,<br>791, 885, 1123, 1215, 3507 |
| TS(HCO →H + CO) | -130.7 | C   0.00000000  -0.20785418  -0.56329970<br>O   0.00000000   0.08229406   0.53562083<br>H   0.00000000   1.17058649  -1.78960978 | 827i,<br>394, 2091 |
| TS(HOC →H + CO) | -5.6 | O   0.00000000  -0.09023806  -0.46403336<br>C   0.00000000   0.04627093   0.71920704<br>H   0.00000000   0.88099954  -1.20457608 | 3142i,<br>1029, 1815 |